\documentclass[aps,prl,groupedaddress,superscriptaddress,twocolumn,notitlepage,bibnotes]{revtex4-1}
\usepackage{graphicx}
\usepackage{hyperref}
\usepackage{physics}
\usepackage{color}
\usepackage{enumerate}
\usepackage{bm}
\usepackage[normalem]{ulem}
\usepackage{comment}

\usepackage[caption=false]{subfig}

\usepackage{graphicx}
\usepackage{braket}
\usepackage{amsmath}

\usepackage{amsthm}
\usepackage{amsfonts, amssymb}
\usepackage{bbm}
\usepackage{hyperref}

\usepackage{mathtools}
\usepackage{mathrsfs}

\DeclarePairedDelimiterX{\infdivx}[2]{(}{)}{%
	#1\;\delimsize\|\;#2%
}

\begin{document}

\newtheorem{theorem}{Theorem}
\newtheorem{proposition}{Proposition}
\newtheorem{corollary}{Corollary}

\theoremstyle{definition}
\newtheorem{definition}{Definition}[section]
\newtheorem{assumptions}{Assumptions}[section]
\newtheorem{example}{Example}[section]
\newtheorem{rmk}{Remark}[section]
\newtheorem{conj}{Conjecture}[section]

\newcommand{\Det}{\text{Det}}
\newcommand{\ergo}{\mathcal{E}}
\newcommand{\ergoparam}{\widetilde{\mathcal{E}}}
\newcommand{\en}{\mathfrak{E}}
\newcommand{\freeen}{\mathcal{F}}
\newcommand{\pass}{^{\downarrow}}
\newcommand{\dstate}{\hat{\rho}}
\newcommand{\cohstate}{\hat{\varphi}}
\newcommand{\optstate}{\hat{\rho}^\ast}
\newcommand{\sstate}{\hat{\xi}}
\newcommand{\thermstate}{\hat{\tau}_{\beta}}
\newcommand{\ham}{\hat{H}}
\newcommand{\hilb}{\mathcal{H}}
\newcommand{\denspace}{\mathcal{D}(\mathcal{H})}
\newcommand{\cohspace}{\mathcal{C}(\mathcal{H})}
\newcommand{\pos}{\hat{q}}
\newcommand{\mom}{\hat{p}}
\newcommand{\destr}{\hat{a}}
\newcommand{\constr}{\hat{a}^{\dagger}}
\newcommand{\fixspace}{\mathcal{D}(\mathcal{H},E)}
\newcommand{\fixcoh}{\mathcal{C}(\mathcal{H},E)}
\newcommand{\erre}{\hat{r}}
\newcommand{\weylop}{\hat{D}}
\newcommand{\unitrasf}{\hat{V}}
\newcommand{\dagg}{^{\dagger}}
\newcommand{\gstate}{\hat{\rho}}
\newcommand{\enopt}{ENOPT_{\Phi,E}}
\newcommand{\eropt}{EROPT_{\Phi,E}}
\newcommand{\amp}{\mathcal{A}}
\newcommand{\loss}{\mathcal{L}}
\newcommand{\noise}{\mathcal{N}_{N}}

\newcommand{\sigmax}{{\bf{\sigma}_x}}
\newcommand{\sigmay}{{\bf{\sigma}_y}}
\newcommand{\sigmaz}{{\bf{\sigma}_z}}

\title{Quantum  Energy  Lines and the optimal output ergotropy problem}
\author{Salvatore Tirone}
	\email[]{salvatore.tirone@sns.it}
\affiliation{Scuola Normale Superiore, I-56127 Pisa, Italy}

\author{Raffaele Salvia}
\email[]{raffaele.salvia@sns.it}
\affiliation{Scuola Normale Superiore, I-56127 Pisa, Italy}

\author{Vittorio Giovannetti}
\affiliation{NEST, Scuola Normale Superiore and Istituto Nanoscienze-CNR, I-56127 Pisa, Italy}

\begin{abstract}
We study the transferring of useful energy (work) along a transmission line that allows for partial preservation of quantum coherence. As a figure of merit we adopt the maximum values that ergotropy, total ergotropy, and non-equilibrium free-energy attain at the output of the line for an assigned input energy threshold. For Phase-Invariant Bosonic Gaussian Channels (BGCs) models, we show that coherent inputs are optimal. For (one-mode) not Phase-Invariant BGCs we solve the optimization problem under the extra restriction of Gaussian input signals.
\end{abstract}

	\maketitle

	\textbf{\em Introduction}:-- 
	 Quantum technologies,  which are extremely successful in delivering groundbreaking improvements for information processing procedures~\cite{Dowling},
	 have a chance of being essential also in the management of energy sources.
In particular 
the tremendous advances in experimental techniques witnessed  in the last decade~\cite{THERM1,THERM2,THERM3,THERM4,THERM5, Peterson2019}  suggest the possibility of realistically
enhancing the performances 
of thermal machines by designing new types of devices  that 
maintain some degree of quantum coherence in their functioning (quantum thermal machines) 
~\cite{NEW1,NEW2,NEW3, NEW4,NEW5,NEW6,NEW7,NEW8,NEW9,NEW10,NEW11,NEW12,NEW13,NEW14,NEW15,GelbwaserKlimovsky2013, GelbwaserKlimovsky2014, GelbwaserKlimovsky2015, Friis2018, PerarnauLlobet2015b, PhysRevE.91.032119, Mukherjee2016, PhysRevA.99.052320, Francica2020,  Camati2019, Francica2019, Latune2019, Monsel2020, Alicki2013}.
	Furthermore recent studies~\cite{Campaioli2018, Andolina2019, Farina2019, Rossini2019, rossini2019quantum, Rosa2020}  indicate that  using genuine quantum systems as	energy storing devices (quantum batteries)  
  could  be crucial in speeding up energy charging processes -- see Ref.~\cite{QUACH2021} for a first  experimental proof-of-concept implementation. 
In view of these results, it 
 makes sense to study the impact that quantum effects  may have 
on  energy transmission procedures.
  Here we present a first study of Quantum Energy Lines (QELs in brief) which, at variance with traditional models, are capable to preserve a certain degree of quantum coherence
 during the transfer of energy  pulses.
Previous works on the subject can be found in  quantum biology, 
where it was observed the rather counter-intuitive feature
  of  noise-enhanced speedup~\cite{Rebentrost_2009, Biggerstaff2016,  Uchiyama2018} which may actually contribuite to the efficiency of the light-harvesting complexes responsible of photosintesis~\cite{Plenio_2008, Rebentrost2009, Mohseni2013, Jeske2015,  Levi2012}. Moreover, there have been theoretical attempts to teleport energy using ground state fluctuations of quantum fields~\cite{Hotta2008, Ye2016}. 
  At variance with those studies, where the goal of the transmission line is to send down energy from {\it classical} energy sources to {\it classical} users,
  in our vision QELs could be employed 
to improve the connectivity  between 
 energy power plants and energy storing sites
that are capable to handle energy in a quantum coherent
 fashion, avoiding the
need to pass through unnecessary quantum-to-classical and classical-to-quantum conversion stages~-- see e.g.~\cite{maffei2021probing}.

  While rather unconventional, prototypical examples of QELs already exist in the form of 
  free space or optical fiber transmission lines which are 
currently under development by various international agencies~\cite{DARPA1,DARPA2,SECOQC,XU,SWISS,TOKYO}. These  schemes have been extensively studied
by the quantum information 
 community~\cite{CAVES,BOSrew1,Serafini2017,Holevo2012,Wilde_2013,PhysRevResearch.3.013279,cariolaro,PhysRevLett.92.027902,PhysRevA.68.062323,PhysRevLett.98.130501}  and admit a formal description  in terms 
  the formalism of Bosonic Gaussian Channels (BGCs)~\cite{CAVES,BOSrew1,Serafini2017,Holevo2012,Wilde_2013} 
  which we shall adopt hereafter.  
 Within this setting, the main goal of our analysis is to identify the
 pulses that have to be sent through the QEL to ensure the lowest level of energy waste. Such
 task is not dissimilar from the optimization problem one faces with more conventional 
 power lines: in the present case however the issue is  complicated by the absence of a clear 
 distinction between heat and work  in a  purely quantum mechanical setting~\cite{ Scovil59,Alicki79,Kosloff84, Alipour2016}.
Singling out the useful part (work) of the internal energy of a quantum system
 (in our case the output signals of the QEL), 
 does indeed strongly depend upon the resources available to the process~\cite{Niedenzu2019}. 
 The maximal amount of energy one can recover by means of reversible coherent (i.e. unitary)
 processes~\cite{def_ergo} is the \textit{ergotropy}~$\ergo$, a non-linear functional of the state of the quantum system~\cite{def_ergo}.
 This quantity is relevant in the study of cycles~\cite{GelbwaserKlimovsky2013, GelbwaserKlimovsky2014, GelbwaserKlimovsky2015, Friis2018, PerarnauLlobet2015b, PhysRevE.91.032119} and it has important connections with the theories of entanglement and coherence~\cite{Mukherjee2016, PhysRevA.99.052320, Francica2020}. 
	If we have instead access to many copies of the system, the relevant quantity is the \emph{total ergotropy}
	$\ergo_{tot}$, which is the work that we reversibly extract from an ensemble of asymptotically many copies of the system of interest~\cite{PhysRevA.96.052112, Niedenzu2019}.  Finally, granting access to a thermal bath, we can push the work extraction process to a further level
quantified by the {\it non-equilibrium free-energy} $\freeen^\beta$ of the state~\cite{Esposito2011, Brando2015, Niedenzu2016}.

Ergotropy, total ergotropy, and non-equilibrium 
free-energy all qualify as bona-fide figures-of-merit for the work we can extract from a quantum system. 
 Accordingly we shall study the efficiency of a QEL by determining  which, among the set
of   signals that have the  same input energy, ensure the highest values 
of $\ergo$,$\ergo_{tot}$, and $\freeen^\beta$ at the receiving end of the line.  
For the special case of QELs described by Phase-Insensitive (PI) BGCs which describe propagation loss, thermalization and amplification noise effects,  we provide an exact solution of the optimization task
showing that optical coherent states~\cite{mandel_wolf_1995, scully1997quantum} always ensure the best performances for all the three figures-of-merit (a result that mimics the Gaussian Optimization solution observed 
in the study of PI-BGC as quantum communication lines~\cite{mari2014,maj_multi,SOL1,SOL2}). We complete our analysis presenting the solution for the optimization problem for the special case of general (non-PI) one-mode BGCs, obtained limiting the input signals to
Gaussian states~\cite{Serafini2017}.

		\textbf{\em The scheme}:-- We shall model a QEL as a collection of Bosonic (electromagnetic) modes  that lose energy en route from the transmitter to the receiver while possibly undergoing events of amplification, and thermalization effects~\cite{BOSrew1,PhysRevResearch.3.013279,cariolaro}. 
A rigorous mathematical characterization of the noise affecting the transmitted signals in these lines can be obtained in terms of Phase-Insensitive (PI) BGCs~\cite{Holevo2012,HolevoWerner,Serafini2017,Caruso_2008,caruso_onemode} whose properties we now review in brief. 
An $n$-mode Continuous Variables (CV) system is described by a complex separable Hilbert space $\hilb$ 
	equipped with self-adjoint bosonic field operators $\pos_1, \mom_1, \cdots, \pos_n, \mom_n$   that obey  the canonical commutation relations (i.e. $[\pos_j,\pos_k] = [\mom_j,\mom_k] = 0$,   $[\pos_j,\mom_k] = i \delta_{jk}\hat{I}$ with  $\hat{I}$ the identity operator), and 
by the
	  free  electromagnetic Hamiltonian
	  $\ham := \sum_{j=1}^n\frac{\pos_j^2 + \mom_j^2}{2} - \frac{n}{2}$,
	  which we express  using  dimensionless units ($\hbar = \omega = 1$) and removing the vacuum energy contribution. 
		The set of quantum states $\denspace$ of the system comprises all positive trace-class operators on $\hilb$ with trace~$1$.
	Introducing $\erre := (\pos_1,...,\pos_n,\mom_1,...,\mom_n)^T$  	 
	 and  using $\{,\}$ to denote the anticommutator, for each $\dstate\in \denspace$ we hence define  its statistical mean vector $m(\dstate) := \Tr[\dstate\erre] \in \mathbb{R}^{2n}$, its 
	   covariance matrix $\sigma(\dstate)\in \mathbb{R}^{2n\times 2n}$ of elements 
	$\sigma_{jk}(\dstate) := \Tr[\dstate\{\erre_j - m_j,\erre_k - m_k\}]$, and 
its characteristic function 
	$\chi(\dstate;x) := \Tr[\dstate\weylop(x)]$,
	where $\weylop(x) := \exp[i\erre\cdot x]$ is the Weyl or displacement operator and $x \in \mathbb{R}^{2n}$.
	We also say that $\dstate\in \denspace$ is a Gaussian state  when the associate characteristic function is gaussian~\cite{caruso_onemode, Caruso_2008}, i.e.  when 
	$\chi(\dstate;x) = e^{-\frac{1}{4}x^T\sigma x + im\cdot x}$ for some mean vector $m$ and convariance matrix $\sigma$.
	Given now an input $n_I$-mode CV system with Hilbert space $\hilb_I$, and an output $n_O$-mode system with Hilbert space $\hilb_O$, a BGC $\Phi: \; \mathcal{D}(\mathcal{H}_I) \rightarrow \mathcal{D}(\mathcal{H}_O)$ is a CPTP map~\cite{Holevo2012}  that preserves the gaussian character of the transmitted signals. 
	These transformations can be formally described by
	assigning a vector  $v\in\mathbb{R}^{2n_O}$ and matrices $Y \in \mathbb{R}^{2n_O\times2n_O}$, 
  $X \in \mathbb{R}^{2n_O\times2n_I}$  that verify the condition $Y \geq i(\gamma_{n_O} - X\gamma_{n_I}X^T)$ with  $\gamma_{n}:=
	 \tiny{\begin{pmatrix}
	0 & I_{n} \\
	-I_{n} & 0
	\end{pmatrix}}$ being a  $2n\times 2n$ block matrix.
  Explicitly 
  given $\dstate \in \mathcal{D}(\mathcal{H}_I)$ the state describing the input signal of the channel, the characteristic function
  of the corresponding state  $\Phi(\dstate)$ at the end of the transmission line can be written as 
	$\chi(\Phi(\dstate);x) = \chi(\dstate;X^Tx)e^{-\frac{1}{4}x^TYx + iv\cdot x}$, implying the identities 
	\begin{equation} \label{gauss_action}
	 m((\Phi(\dstate)) = Xm(\dstate) + v\;, \quad \sigma((\Phi(\dstate)) =X\sigma(\dstate) X^T + Y\;.
	\end{equation}
	  A BGC map $\Phi$  is finally said to be  Phase-Insensitive (PI) if for all input states $\dstate\in \mathcal{D}(\mathcal{H}_I)$, and for all $t \in \mathbb{R}$ 
	  we have 
	\begin{eqnarray} \label{def:gauge_cov}
	\Phi(e^{-i\ham_I t}\dstate e^{i\ham_I t}) = e^{\mp i\ham_O t}\Phi(\dstate)e^{\pm i\ham_O t},
	\end{eqnarray}
	with $\ham_I$ and $\ham_O$ the free Hamiltonians of $\hilb_I$ and $\hilb_O$ respectively (specifically a $\Phi$ 
	verifying~(\ref{def:gauge_cov}) with the upper signs in the r.h.s. is  said to be \emph{gauge-covariant}). 
	From a practical point of view
	the multimode 
	PI-BGCs models described here provide an idealized yet commonly used version of broadband communication lines, 
	 because they characterize quantum states transferred through an optical medium via electromagnetic pulses whose bandwdith is small compared to their central reference frequency~\cite{CAVES,PhysRevLett.92.027902,PhysRevA.68.062323,PhysRevLett.98.130501}.

	\textbf{\em Quantum Work-Extraction}:-- 
At variance with purely classical settings,   
 discriminating which part of the internal energy 
of a quantum system (e.g. the output signal of a QEL) can be identified with
 heat or work is 
difficult~\cite{Scovil59,Alicki79,Kosloff84} due to 
correlation-induced  entropy increases that may arise when coupling the system to an external {load}~\cite{Niedenzu2019}.
Nonetheless, limiting the allowed  operations to be local, fully reversible, and coherent (i.e. unitary), 
the amount of work we can extract from a
 single copy of a density matrix $\dstate$ of a system
is given by the \emph{ergotropy} functional $\ergo(\dstate)$~\cite{def_ergo}.
Letting $\en(\dstate):=\Tr[ \dstate \ham]$, 
we can write
	\begin{equation} \label{def:ergo}
		 \ergo(\dstate) :=
		 \en(\dstate) - \en(\dstate\pass) \;, 
		\end{equation}
	where $\dstate\pass$ is the {\it passive counterpart}~\cite{Pusz1978, Lenard1978} of $\dstate$, i.e. the special element of $\denspace$  which has the lowest energy among those with the same spectrum of~$\dstate$~\cite{SM}. 
Since passive states are not necessarily {\it completely passive}~\cite{skrzypczyk2015, Salvia2020energyupperbound},
	ergotropy  turns out to be a non-extensive, super-additive quantity.
	 Accordingly, when operating with reversible coherent operations on  $N$ copies of a given state $\dstate$, 
	 it is possible  to increase the total amount of extractable energy by acting jointly on the whole set of subsystems. 
	 The maximum amount of energy per copy that is attainable under this new paradigm is quantified by the {\it total ergotropy} $\ergo_{tot}(\dstate)$, a functional fulfilling the inequality $\ergo_{tot}(\dstate) \geq \ergo(\dstate)$ which can obtained
	via a proper  regularization of~(\ref{def:ergo}), i.e. 
	\begin{eqnarray}
	\label{ergo_tot}
	\ergo_{tot}(\dstate) := \lim_{n \to \infty} \frac{1}{n} \ergo(\dstate^{\otimes n}) = 
	\en(\dstate) - \en(\hat\tau_{\beta(\dstate)}) \; ,
	\end{eqnarray}
	where in the last identity  $\thermstate := {e^{-\beta \ham}}/{\Tr[e^{-\beta \ham}]}$
	 is a thermal Gibbs state of the system  whose inverse temperature $\beta \in \mathbb{R}^+$  
	 is fixed in order to ensure  
	$S(\hat\tau_{\beta(\dstate)})=S(\dstate) :=-\Tr[\dstate \log \dstate]$. 
 $\ergo_{tot}(\dstate)$ represents the ultimate amount of energy that we can extract reversibly from $\dstate$ when having at disposal 
 an unlimited number of copies.  
 More energy from the system can still be converted into useful work only if we are willing to admit
 some dissipation side-effect, e.g. by coupling the system with an external thermal bath~\cite{Esposito2011,Niedenzu2019}. In this case the overall amount of extractable energy is provided by the   
 non-equilibrium \emph{free energy} functional:
	$\freeen^\beta(\dstate) := \en(\dstate) - S(\dstate)/\beta$,
	with $\beta$ representing the inverse temperature of the bath.

	\textbf{\em Optimal inputs for PI-BGCs}:--
	Here we present our main result: input coherent states \cite{Serafini2017} maximize the three functionals introduced in the previous sections at the output of any PIBGC. To this aim we first observe the following fact: 
	 	\begin{theorem}
			Given   $\Phi$  a  PI-BGC  from $n_I$ input  to $n_O$ output modes,
		for any 
		$\dstate \in \mathcal{D}(\hilb_I)$
		there exists at least a coherent input state $\cohstate\in \mathcal{D}(\hilb_I)$ having the same mean input energy of  $\dstate$ 
				 such that $\en(\Phi(\cohstate)) \geq \en(\Phi(\dstate))$. 
		\label{lemma:max_en}
	\end{theorem}
	\begin{proof}
			We recall that 
			  the mean energy of a quantum state $\dstate\in \mathcal{D}(\hilb)$ of $n$ modes can be expressed in terms of its statistical mean and covariance matrix via the compact expression 
		$\en(\dstate) = \frac{\Tr[\sigma(\dstate)]}{4} + \frac{|m(\dstate)|^2}{2}  - n/2$. 
			Recall also that the coherent states $\cohstate$ of
					a $n_I$-mode CV system are characterized by a covariance matrix $\sigma(\cohstate) = I_{2n_I}$; so any $\cohstate\in{\mathcal C}({\mathcal H}_I)$ is uniquely identified by its statistical mean~$m$: $\ket{\varphi} = \weylop(m)\ket{\O}$
with $\ket{\O}$ being the vacuum state of the model. 
Thanks to this identity and to Eq.~(\ref{gauss_action}),  the mean energy at  the output of
a BGC $\Phi$ defined by the vector $v$ and the matrices $X$, $Y$ can be expressed as $\en(\Phi(\dstate)) = \frac{\Tr[X^TX \sigma(\dstate)]}{4} + \frac{|Xm(\dstate)+v|^2}{2} + c$,
where $c= \Tr[Y]/4 - n_O/2$ is a constant that is independent from the input state $\hat{\rho}$.
Next we remind that any PI-BGC has $v=0$ and admits an orthogonal, symplectic transformation $V \in  \mathbb{R}^{2n_I\times2n_I}$
such that the following statement holds~\cite{WOLF2008,Serafini2017}: 
\begin{equation}
\label{XTX_diagonale}
\left( V^T X^T X V \right)_{jk} = \Lambda_j \delta_{jk} \; ,
\end{equation}
with 
$\Lambda_{m} = \Lambda_{m+n_I}$  $\forall m=1,...,n_I$, and 
$\Lambda_{m} \geq \Lambda_{m+1}$ $\forall m=1,..,n_I $. 
Observe that the above conditions are equivalent to saying that $\sqrt{\Lambda_1}$ is the highest singular eigenvalue of $X$, or explicitly that for every vector $w \in \mathbb{R}^{2n_I}$ it holds $|Xw|^2 \leq \Lambda_1 w^2$. 
		Now consider a coherent state $\cohstate$ with mean vector $m(\cohstate)$ 
		oriented in such a way to
		 saturate the former inequality, i.e. 
		$|Xm(\cohstate)|^2 = \Lambda_1 m(\cohstate)^2$,
		that is  with a $m(\cohstate)$ which is an eigenvector of the matrix $X^TX$. Notice that such condition can be fulfilled for any value of $\lvert m(\cohstate) \rvert$, and hence for any possible $\en(\cohstate)$.
		For such a choice we can hence write  the inequality 
		\begin{equation} 
	{|Xm(\cohstate)|^2 - |Xm(\dstate)|^2}
				\geq 
		\Lambda_1 ( m(\cohstate)^2 - m(\dstate)^2 )\;,
		\label{ergodiff_latom}
		\end{equation}
		that holds true for all $\dstate \in \mathcal{D}(\hilb_I)$.
		Notice also that thanks to~(\ref{XTX_diagonale}) and remembering that $\sigma(\cohstate) = I_{2n_I}$, we have		\begin{eqnarray} 
		&&{\Tr[X^TX (\sigma(\dstate)-\sigma(\cohstate))]} = 
		\sum_{j=1}^{2n_I} \Lambda_j \left[ (V\sigma(\dstate)V^T)_{jj} - 1 \right] \nonumber
		\\ &&   \leq
		\Lambda_1 {\Tr[V\sigma(\dstate)V^T - \sigma(\cohstate)]}=\Lambda_1 {\Tr[\sigma(\dstate) -\sigma(\cohstate)]} 		\label{ergodiff_latosigma}\;. 
		\end{eqnarray} 
		In deriving~\ref{ergodiff_latosigma} we exploited the following two facts: i) since  $V$ is a symplectic matrix, then $V\sigma(\dstate)V^T$ is a covariant matrix $\sigma(\dstate')$ of a proper
		density matrix $\hat{\rho}'$ of the system;
		ii) for all covariant matrices $\sigma$  it is always true that  
		$\sigma_{m,m} + \sigma_{m+n_I,m+n_I} \geq 2$ for all $m=1,\cdots n_I$,
		 which can be easily proven by noticing that the left-hand side is the trace of the covariance matrix of the reduced density matrix of the $m$-th mode of the input system
		 and from the fact that  for any $n$-mode 
		 quantum state $\Tr[\sigma(\dstate)] \geq 2n$, since $\en(\dstate) \geq 0$. 
		Finally, using~(\ref{ergodiff_latom}) and~(\ref{ergodiff_latosigma}) 
		we can conclude that
		\begin{eqnarray} 
		&&\en(\Phi(\cohstate)) - \en(\Phi(\dstate))\\  \nonumber  && = 
		\tfrac{|Xm(\cohstate)|^2 - |Xm(\dstate)|^2}{2} - \nonumber \tfrac{\Tr[X^TX (\sigma(\dstate)-\sigma(\cohstate))]}{4}  \\&&  \geq 
		\Lambda_1 \left[ \tfrac{  m(\cohstate)^2 - m(\dstate)^2  }{2}  - \tfrac{\Tr[\sigma(\dstate)-\sigma(\cohstate)]}{4} \right] = \Lambda_1 \left[ \en(\cohstate) - \en(\dstate) \right] \geq 0 \; ,
		\nonumber
		\end{eqnarray}
		which evaluated in the case where $\cohstate$ and $\dstate$ shares the same input energy  (i.e. $\en(\cohstate)=\en(\dstate)$) implies $\en(\Phi(\cohstate)) \geq  \en(\Phi(\dstate))$, hence  the thesis. 
		\end{proof}
		Exploiting the above result we are now ready to present our main finding: 
	\begin{theorem}
	\label{theorem1} Given  $\Phi$ a PI-BGC from $n_I$ input  to $n_O$ output modes, for any 
	$E \in \mathbb{R}^+$ and $\dstate\in  \mathcal{D}(\hilb_I)$ with $\en(\dstate) \leq E$, 
	there exists  a coherent state $\cohstate \in \mathcal{D}(\hilb_I)$ with $\en(\cohstate) = E$ that achieves higher values of the output
	ergotropy, max ergotropy, and non-equilibrium free-energy functionals, i.e. 
	\begin{eqnarray}
	&&\ergo(\Phi(\cohstate)) \geq \ergo(\Phi(\dstate)) \;, \quad  
	\ergo_{tot}(\Phi(\cohstate)) \geq \ergo_{tot}(\Phi(\dstate)) \nonumber \;, \\
&& \freeen^\beta(\Phi(\cohstate)) \geq \freeen^\beta(\Phi(\dstate)) \;,   \quad \forall \beta > 0  \; . \label{RESULT_freeen}
	\end{eqnarray}	
	\end{theorem}
	
	\begin{proof}	
	The ergotropy, the total ergotropy, and the non-equilibrium free energy are all Schur convex functional of the states (see~\cite{SM} for details, which includes Refs. \cite{MarshallOlkin,Alimuddin2020,nielsen_chuang_2010,Li2013,Shirokov2020}).
	Now in Refs.~\cite{mari2014,maj_multi} it has been shown that coherent states optimize the output of BGCs with respect to every Schur-convex functional.  Therefore, 
		given $\en(\dstate) \leq E$ and  every coherent state $\cohstate$ we can write
	\begin{eqnarray}
&&\en(\Phi(\cohstate)\pass) \leq \en(\Phi(\dstate)\pass) \;  , \quad \en(\hat\tau_{\beta(\Phi(\cohstate))}) \leq \en(\hat\tau_{\beta(\Phi(\dstate))})\;, \nonumber 
\\	
&& S(\Phi(\cohstate)) \leq S(\Phi(\dstate)) \;.  
	 \label{lemma3_Eth}
	\end{eqnarray}
	The thesis now immediately follows from the above expressions and from  \textbf{Theorem \ref{lemma:max_en}} which guarantees that there is at least a coherent state $\cohstate$ with mean energy greater or equal to $E$
 that fulfils the inequality $\en(\Phi(\cohstate)) \geq \en(\Phi(\dstate))$.
 	\end{proof}

	\textbf{\em One-mode PI-BGCs}:--
    In the special case of one-mode (i.e. $n_I=n_O=1$) PI-BGCs, some simplification occurs that allows
	us to extend a little further the result of the previous section. 
	First we remark that in this context 
	passivity and complete passivity are equivalent notions\cite{Serafini2017}:
	\begin{eqnarray}
	\label{ergotot_unmodo}
	\ergo(\hat{\tau}) = \ergo_{tot}(\hat{\tau}) \quad \text{$\forall \hat{\tau}$ one-mode Gaussian states.}
	\end{eqnarray} 
	In this particular setting the energy at the output of PI-BGCs $\mathfrak{E}[\Phi(\rho)]$ depends only on the input energy $\mathfrak{E}(\rho)$, using both this fact and the main result of \cite{DePalma2016}, one can prove the following statement:
	\begin{theorem} Given $\Phi$  a one-mode PI-BGC and 
	for any $E,s \in \mathbb{R}^+$ and one-mode bosonic state $\dstate$ with $\en(\dstate) \leq E$ and $S(\dstate) \geq s$, there exists a gaussian state $\hat{\tau}$ with mean energy $E$ and entropy $s$,
	 that achieves higher values of the output
	ergotropy, max ergotropy, and non-equilibrium free-energy functionals, i.e. 
	\begin{eqnarray}
	&&\ergo(\Phi(\hat{\tau})) =\ergo_{tot}(\Phi(\hat{\tau}))  \geq \ergo_{tot}(\Phi(\dstate)) \geq \ergo(\Phi(\dstate)) \;, 
\nonumber \\
&& \freeen^\beta(\Phi(\hat{\tau})) \geq \freeen^\beta(\Phi(\dstate)) \;,   \quad \forall \beta > 0  \; . \label{RESULT_freeen}
	\end{eqnarray}	
\end{theorem}
	It is easy to notice that in this scenario Theorem \ref{theorem1} is a special instance (s=0) of the result above. \\
All one-mode PI-BGCs can be expressed as compositions of three maps~\cite{caruso_onemode}: 
	the lossy thermal channel $\loss_{\eta,N}$, describing the interaction with a thermal environment of mean photon number $N\geq 0$ through 
	a beam-splitter of transmissivity $\eta\in [0,1]$  ($X = \sqrt{\eta} I_2$; $Y = (1-\eta)(2N+1) I_2$); the amplification thermal channel $\amp_{\mu,N}$, describing the interaction with a thermal environment 
	through a linear optical amplifier of gain  $\mu\geq 1$  ($X = \sqrt{\mu} I_2$; $Y = (\mu-1) (2N+1)I_2$);  and the additive classical noise channel $\noise$ describing a random displacement of the signal
	in the phase space  ($X =  I_2$; $Y = 2N I_2$). 
	It follows that, 
	constraining the input energy to be $\en(\dstate) \leq E$,  the 
	  maximum  ergotropy (and total ergotropy) achievable at the output	are respectively 
$\ergo^{(\max)}_E(\loss_{\eta,N}) = \eta E$, $\ergo^{(\max)}_E(\amp_{\mu,N}) = \mu E$, and 
	$\ergo^{(\max)}_E(\noise) = E$ -- see~\cite{SM} for details. Notice that the reported  values do not depend upon $N$ and exhibit the multiplicative behaviour found in~\cite{tirone2021kelly}, where an optimization of the output ergotropy for $\loss_{\eta,N}$ and $\amp_{\mu,N}$  was performed on the restricted setting of Gaussian inputs.

		\textbf{\em BGCs which are not PI}:--
	 If we drop the phase invariance assumption~(\ref{def:gauge_cov}), coherent states $\cohstate$ no longer represent the optimal choices for the output work-extraction functionals: in this case the problem is made more complex by the fact that now 
 the channel does not admit a single input state $\dstate$ that maximizes the positive contribution  of
	$\ergo(\Phi(\dstate))$, $\ergo_{tot}(\Phi(\dstate))$, and
	  $\freeen^\beta(\Phi(\dstate))$ (i.e. the term $\en(\Phi(\dstate))$), and at the same time minimizes the negative one (e.g. $\en(\Phi(\dstate)\pass)$ for the ergotropy). One-mode not phase invariant  BGC channel are able to describe energy exchanges of the transmitted signals with a squeezed vacuum environment\cite{squeezedvac1, squeezedvac2}. In\cite{SM} we consider in full generality one-mode BGCs. To elucidate the difficulty of the problem, here we show the example of the map $\Gamma_{\eta,\zeta} := \loss_{\eta,0} \circ \Sigma_{\zeta}$, results from a concatenation of  a squeezing unitary evolution ($X = \tiny{\begin{pmatrix} \sqrt{\zeta} & 0 \\ 0  & \sqrt{\zeta} \end{pmatrix}}$; $Y = 0$) that precedes the action of a 
quantum-limited attenuator  (here $\zeta \geq 1$, with $\zeta=1$ corresponding to the zero-squeezing case).
	On one hand, for this channel the pure displaced-squeezed states $\gstate_1$ having covariance matrix $\sigma(\gstate_1) = \tiny{\begin{pmatrix} \zeta^{-1} & 0 \\ 0 & \zeta \end{pmatrix}}$ can be easily shown to provide the output configurations that majorizes all the others~\cite{NielsenVidal2001}, minimizing the negative contributions of $\ergo(\Gamma_{\eta,\zeta}(\dstate))$, $\ergo_{tot}(\Gamma_{\eta,\zeta}(\dstate))$, and
	  $\freeen^\beta(\Gamma_{\eta,\zeta}(\dstate))$ by the same Schur-convex argument we used before -- 
		indeed, with this choice $\Sigma_{\zeta}(\gstate_1)$ becomes a coherent state, and the result
	follows directly from Refs.~\cite{maj_multi,mari2014} by observing that  $\loss_{\eta,0}$ is phase-insensitive.
	On the other hand, let $\gstate_2$ be the gaussian state with moments $m(\gstate_2) = m(\gstate_1)$ and $\sigma(\gstate_2) = \tiny{\begin{pmatrix} \zeta & 0 \\ 0 & \zeta^{-1} \end{pmatrix}}$. It is not difficult to see that $\gstate_2$ has the same energy as $\gstate_1$, but that $\en(\Gamma_{\eta,\zeta}(\gstate_2)) > \en(\Gamma_{\eta,\zeta}(\gstate_1))$ for all $\zeta\geq 1$,  preventing  $\gstate_1$ from being
	the optimal choice for the positive contribution of the output work extraction functionals.

\begin{figure}
	\centering
	\includegraphics[width=\columnwidth]{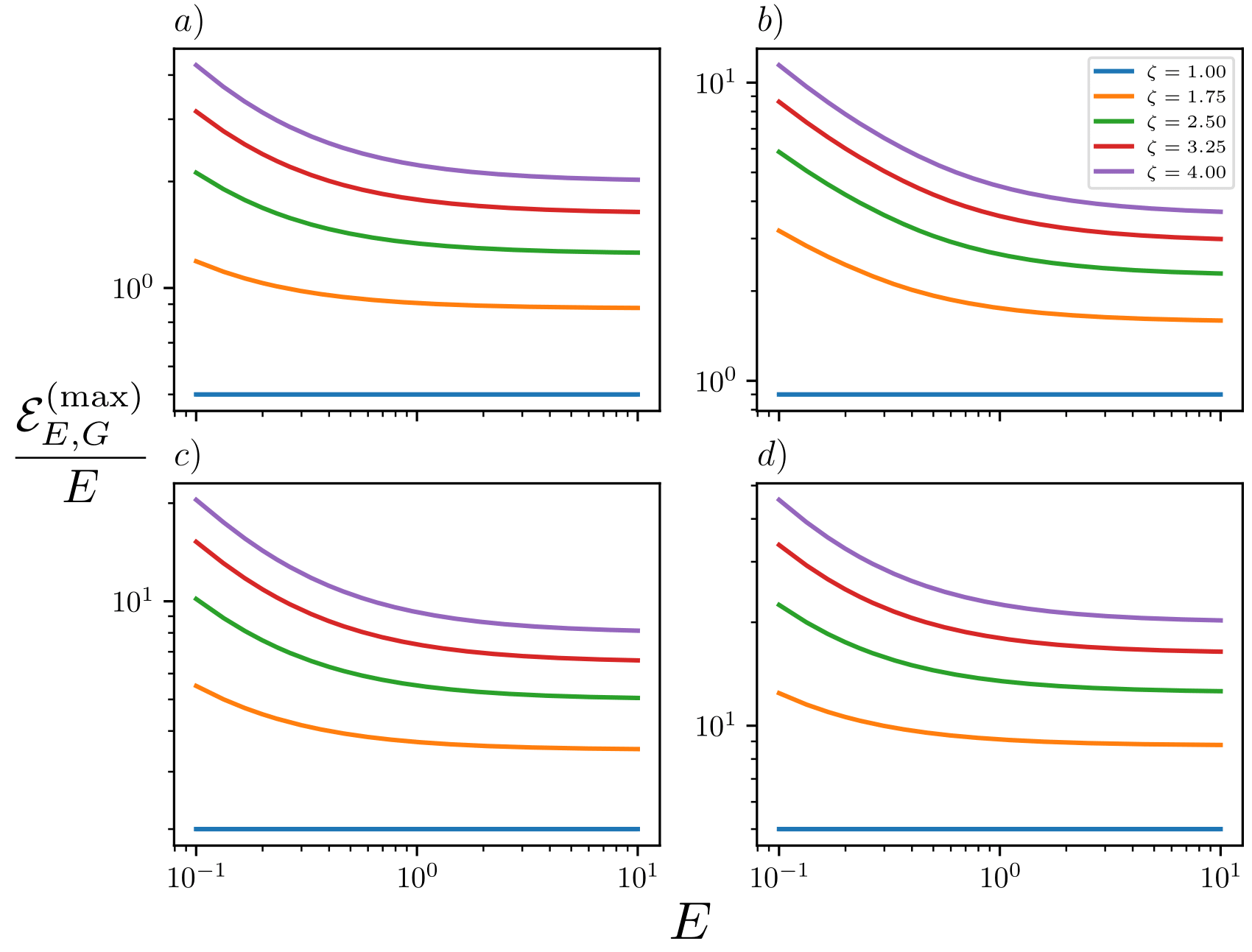}
	\caption{Rescaled maximum output ergotropy values $\ergo^{(\max)}_{E,G}/E$ attainable with Gaussian inputs with input energy $E$ for  one-mode, 
	not-PI BGCs. {\it a)} and {\it b)}  attenuator-squeezing channels  $\Gamma_{\eta,\zeta}=\loss_{\eta,0} \circ \Sigma_{\zeta}$  with $\eta =0.5$ and $\eta=0.9$
	respectively;  {\it c)} and {\it d)} amplifier-sqeezing channels $\Theta_{\mu,\zeta}=\amp_{\mu,0}\circ \Sigma_{\zeta}$ with $\mu=2$ and $\mu=5$ respectively. 
In the no-squeezing  $\zeta=1$ regime (blue lines)  the maps are PI and the reported  values   coincide with the absolute maxima ($\eta$ and $\mu$)
 dictated by Theorem \ref{theorem1}.  
 }
	\label{fig:ErgoRatio}
\end{figure}

		A partial solution of the optimal work preservation problem  at the output of non-PI BGCs is presented in Ref.~\cite{SM} where, focusing on one-mode (not-PI) BGCs, we
		provide an  analytical characterization  of the maximal output ergotropy $\ergo^{(\max)}_{E,G}$ achievable using energy constrained Gaussian inputs
		(incidentally, thanks to (\ref{ergotot_unmodo}) these values also coincide with the Gaussian maximal values of the output total ergotropy). 
		The results we obtained are summarised in Fig.~\ref{fig:ErgoRatio} where we plot the ratio  $\ergo^{(\max)}_{E,G}/E$ for different types of channels 
		$\Gamma_{\eta,\zeta}=\loss_{\eta,0} \circ \Sigma_{\zeta}$ and $\Theta_{\mu,\zeta}=\amp_{\mu,0}\circ \Sigma_{\zeta}$ obtained respectively by composing attenuator and amplifying channels with squeezing operations. Notice that the presence of squeezing tends to boost the  ergotropy throughput by yielding values of the ratio which can 
		exceed 1 even in the presence of attenuation, and that in the high energy limit  the  solutions approach the asymptotic limits
		 $\lim_{E\rightarrow\infty} \ergo^{(\max)}_{E,G}(\Gamma_{\eta,\zeta})/E =\eta\zeta$ and $\lim_{E\rightarrow\infty}  \ergo^{(\max)}_{E,G}(\Theta_{\mu,\zeta})/E = \mu\zeta$, respectively.

\textbf{\em Conclusions}:--
The study of QELs paves the way to design improvements  for quantum batteries
or quantum thermal engines by facilitating the interconnections between cluster of such devices, and contributing to the stabilizing protocols for preserving the energy stored within~\cite{Liu2019, PhysRevE.100.032107, PhysRevApplied.14.024092, PhysRevA.102.060201, PhysRevE.101.062114, PhysRevResearch.2.013095, mitchison2020charging, liu2021boosting, santos2020quantum}.
Generalization of the present approach to finite-dimensional setting may represent an interesting
theoretical research line. 
	
\begin{acknowledgments}
We thank G. M. Andolina,  F. Belliardo, and M. Fanizza for insightful discussions.
We acknowledge support by MIUR via PRIN 2017 (Progetto di Ricerca di Interesse Nazionale): project QUSHIP (2017SRNBRK).
\end{acknowledgments}

\end{document}


\maketitle

 The Supplemental material is organized as follows: in Sec.~\ref{sec1} we review the notion of passive and completely passive states; in Sec.~\ref{sec2} 
we discuss Shur convexity and convexity properties of our figure of merit; in Sec.~\ref{sec3} instead we give a detailed account of the optimization for the work extraction 
at the output of one-mode BGCs. 

\section{Passive and Completely Passive states}\label{sec1} 
Here we clarify the notion of passive counterpart of quantum state which enters in the definition of the ergotropy and the {total ergotropy} (a quantity that is sometime called
\emph{asymptotic ergotropy}~\cite{Farina2019}) given in  Eqs.~(3) and (4) of the main text. 

By the spectral theorem we can decompose any (non necessarily passive) state of the system as 
$\dstate = \sum_{k \in \mathbb{N}}p_k\ket{\Phi_k}\bra{\Phi_k}$, where the set $\{\ket{\Phi_k}\}_{k\in\mathbb{N}}$ is an orthonormal basis on the space $\hilb$, and $p_k$ are probabilities  which we can  list in decreasing  order $p_{k+1} \leq p_k$ without loss of generality.
The passive counterpart of $\dstate$ is hence defined as the density matrix 
\begin{eqnarray} \label{def:pass_st}
\dstate\pass = \sum_{k\in\mathbb{N}}p_k\ket{F_k}\bra{F_k},
\end{eqnarray}
where $\ket{F_k}$ are the Fock eigenstates of $\hat{H}$  ordered such that $\ham\ket{F_k} = \varepsilon_k\ket{F_k}$ and $\varepsilon_{k+1} \geq \varepsilon_k$.
Since $\dstate$ and $\dstate\pass$ share the same spectrum it is clear that they are related by a unitary mapping, in particular we can write 
\begin{eqnarray}\label{unitaryeq} 
\dstate\pass = \unitrasf \dstate\unitrasf \dagg\;,\qquad 
\unitrasf  :=  \sum_{k\in\mathbb{N}} \ket{F_k}\bra{\Phi_k}\;. 
\end{eqnarray} 

Passive  states are such that their mean energy   can not decrease after any  transformation
$\hat{V}$ of the unitary set $U(\hilb)$~\cite{Pusz1978, Lenard1978}, i.e. \begin{eqnarray}
\en(\dstate\pass) \leq \en(\hat{V}\dstate\pass\hat{V}^{\dagger}), \label{NEWPROP1} 
\end{eqnarray}
where $\en(\dstate)= \mbox{Tr}[ \hat{H} \dstate]$ is the mean energy functional of the model.

In general, the state $\dstate\pass\otimes\dstate\pass$ is not a passive state, so we can 
introduce the condition of complete passivity~\cite{skrzypczyk2015}: a state is completely passive if $\forall k \in \mathbb{N}$ $(\dstate\pass)^{\otimes k}$ is a passive state. It can be proven \cite{Lenard1978, skrzypczyk2015} that a state  whose support is not entirely contained in the ground state of the Hamiltonian is completely passive if and only if it is a Gibbs state $\thermstate$, that is if	
\begin{eqnarray}
\label{Gibbs_states}
\thermstate := \frac{\exp[-\beta \ham]}{\Tr[\exp[-\beta \ham]]} \; ,
\end{eqnarray}
for some $\beta \in \mathbb{R}^+$. Accordingly  the completely passive counterpart of a quantum state $\dstate$ corresponds to 
the Gibbs state  $\hat\tau_{\beta(\dstate)}$ which share the same von Neumann entropy with the state $\dstate$, i.e. 
\begin{eqnarray} S(\hat\tau_{\beta(\dstate)})=S(\dstate)\;. \end{eqnarray}

\section{Shur convexity and convexity properties of the work extraction functionals}\label{sec2} 

Now we recall the concept of \textit{majorization} \cite{MarshallOlkin, NielsenVidal2001}. Given two states $\dstate = \sum_{k\in\mathbb{N}}p_k\ket{\Phi_k}\bra{\Phi_k}$ and $\sstate = \sum_{k\in\mathbb{N}}q_k\ket{\phi_k}\bra{\phi_k}$ with their eigenvalues arranged in decreasing order, we say that $\dstate$ majorizes $\sstate$ if 
\begin{eqnarray}
	\dstate \succ \sstate \; \Leftrightarrow \; \forall k\in\mathbb{N} \; \sum_{j=0}^kp_j \geq \sum_{j=0}^kq_j, \; .
\end{eqnarray}
Given a function $f$ : $\denspace \rightarrow \mathbb{R}$ we say that $f$ is Schur-convex if 
\begin{eqnarray} \label{def:schur_conv}
	\dstate \succ \sstate \Rightarrow f(\dstate) \geq f(\sstate),
\end{eqnarray}
on the contrary we call $f$ Schur-concave if
\begin{eqnarray} \label{def:schur_conc}
	\dstate \succ \sstate \Rightarrow f(\dstate) \leq f(\sstate) \; .
\end{eqnarray}

The output of any Schur-convex functional is optimised by coherent states:
\begin{lemma}
	Let $f(\dstate)$ be any Schur-convex function of a quantum state $\dstate$.
\end{lemma}

\begin{lemma}
	The energy $\en(\dstate\pass)$ of the passive state associated to $\dstate$ is a Schur-concave functional of $\dstate$.
\end{lemma}
\begin{proof}
	The proof of this statement can be found in the appendix B of Ref.~\cite{Alimuddin2020}.
\end{proof}

\begin{lemma}
	\label{lemma:entropia-schurconcava}
	The Von Neumann entropy $S(\dstate)$ is a Schur-concave functional of~$\dstate$.
\end{lemma}
\begin{proof}
	This is a classic result for Hilbert spaces of finite dimension \cite{nielsen_chuang_2010}, which in Ref. \cite{Li2013} is generalised to infinite-dimensional Hilbert spaces.
\end{proof}

\begin{lemma}
	The thermal energy $\en(\hat\tau_{\beta(\hat\rho)})$ of the Gibbs state with the same entropy as $\dstate$ is a Schur-concave functional of $\dstate$.
\end{lemma}
\begin{proof}
	It is straightforward to see that this quantity is a monotone in the entropy $S(\dstate)$; the results hence follows from lemma~(\ref{lemma:entropia-schurconcava}).
\end{proof}

\begin{lemma}
	The function ergotropy $\ergo$ is convex on states with finite average energy.
\end{lemma}

\begin{proof} \label{lemma:convexity_ergo}
	Given the states $\dstate_1$, $\dstate_2$ and $\dstate_3 = p\dstate_1 + (1-p)\dstate_2$ with $p \in [0,1]$, invoking~(\ref{unitaryeq})  we find the associated passive states as:
	\begin{eqnarray} \nonumber
	\dstate_1\pass = \unitrasf_1\dstate_1\unitrasf_1\dagg \;, \qquad 
	\dstate_2\pass = \unitrasf_2\dstate_2\unitrasf_2\dagg \;, \qquad 
	\dstate_3\pass = \unitrasf_3\dstate_3\unitrasf_3\dagg,
	\end{eqnarray}
	with $\unitrasf_1$, $\unitrasf_2$, and $\unitrasf_3$ unitaries. 
	Then by invoking the property~(\ref{NEWPROP1}) of passive states we get 
	\begin{eqnarray} 
	&&p\ergo(\dstate_1) + (1-p)\ergo(\dstate_2) - \ergo(p\dstate_1 + (1-p)\dstate_2) \\ \nonumber
	&& \qquad = p[\en(\unitrasf_3\dstate_1\unitrasf_3\dagg) - \en(\unitrasf_1\dstate_1\unitrasf_1\dagg)] + (1-p)[\en(\unitrasf_3\dstate_2\unitrasf_3\dagg) - \en(\unitrasf_2\dstate_2\unitrasf_2\dagg)] \geq 0 \;,
	\end{eqnarray}
which proves the statement.
\end{proof}

\begin{lemma}
	\label{lemma:convexity_ergotot}
	The total ergotropy $\ergo_{\text{tot}}(\dstate)$ is convex on states with finite average energy.
\end{lemma}
\begin{proof}
	By construction, the total ergotropy is the limit of a succession of ergotropy functions. Since the latter are convex by lemma~(\ref{lemma:convexity_ergo}), the total ergotropy is also convex.
\end{proof}

\begin{lemma}
	\label{lemma:convexity_freeen}
	For any $\beta \in (0, \infty)]$, the non-equilibrium free energy $\freeen^\beta(\dstate)$ is convex on states with finite average energy.
\end{lemma}
\begin{proof}
	By the definition $\freeen^\beta(\dstate)$ is the difference between a linear functional and a concave functional. Therefore it is convex.
\end{proof}

\begin{lemma} \label{lemma:pure_opt_ergo}
	For any CPTP map $\Gamma$, among the input state with energy $\en(\dstate) = E$, the states that maximize the output ergotropy $\ergo(\Phi(\dstate))$ are pure states.
\end{lemma}
\begin{proof}
	In Ref.~\cite{Shirokov2020} it is proven that any quantum state $\dstate$ can be expressed as
	\begin{eqnarray} \nonumber
	\dstate = \sum_{k\in\mathbb{N}}p_k\ket{\Phi_k}\bra{\Phi_k} \; ,
	\end{eqnarray}
	where $\ket{\Phi_k}\bra{\Phi_k}$ are pure states with energy $\en(\ket{\Phi_k}\bra{\Phi_k}) = E$, and $\sum_{k=1}^\infty p_k = 1$.
	Using lemma \ref{lemma:convexity_ergo}, we can then write
	\begin{eqnarray} \nonumber
	\ergo(\Gamma(\dstate)) = \ergo\left(\Gamma\left(\sum_{k\in\mathbb{N}}p_k\ket{\Phi_k}\bra{\Phi_k}\right)\right) 
	\leq \sum_{k\in\mathbb{N}}p_k\ergo(\Gamma(\ket{\Phi_k}\bra{\Phi_k})) \; .
	\end{eqnarray}
	Therefore, by definition of mean, we can say that there exists at least one pure state $\ket{\Phi_j}\bra{\Phi_j}$ such that 
	\begin{eqnarray}
	\ergo(\Gamma(\ket{\Phi_j}\bra{\Phi_j})) \geq \sum_{k\in\mathbb{N}}p_k\ergo(\Gamma(\ket{\Phi_k}\bra{\Phi_k}))  = \ergo(\Gamma(\dstate))\; .
	\end{eqnarray}
\end{proof}

\begin{lemma} \label{lemma:pure_opt_ergotot}
	For any CPTP map $\Gamma$, among the input state with energy $\en(\dstate) = E$, the states that maximize the output total ergotropy $\ergo_{\text{tot}}(\Phi(\dstate))$ are pure states.
\end{lemma}
\begin{proof}
	It follows from lemma~(\ref{lemma:convexity_ergotot}), in a completely analogous way as in the proof of lemma~\ref{lemma:pure_opt_ergo}. 
\end{proof}

\begin{lemma} \label{lemma:pure_opt_freeen}
	For any CPTP map $\Gamma$, among the input state with energy $\en(\dstate) = E$, the states that maximize the non-equilibrium free energy $\freeen^\beta(\Phi(\dstate))$ are pure states.
\end{lemma}
\begin{proof}
	It follows from lemma~(\ref{lemma:convexity_freeen}), in a completely analogous way as in the proof of lemma~\ref{lemma:pure_opt_ergo}. 
\end{proof}

\section{One-mode BGCs analysis}\label{sec3} 

We recall that the energy of a Gaussian state can be written as
\begin{eqnarray}
\label{def:energy}
\en(\gstate) = \frac{\Tr[\sigma(\gstate)]}{4} + \frac{m(\gstate)^2}{2}\;, 
\end{eqnarray}
and that  ergotropy of a single-mode Gaussian states can be expressed as~\cite{Serafini2017}
\begin{eqnarray}
\label{ergo_gaussiani_1modo}
\ergo(\gstate) = \frac{\Tr(\sigma(\gstate))}{4} + \frac{m(\gstate)^2}{2} - \frac{\sqrt{\det(\sigma(\gstate))}}{2}.
\end{eqnarray}
The problem of optimising $\ergo(\Phi(\gstate))$ is, therefore, entirely determined by the matrices $X$ and $Y$ of the channel and by the moments $m(\gstate)$ and $\sigma(\gstate)$ of the input state.

\subsection{PI-BGCs}
For simple yet practically useful classes of one-mode PI-BGCs we can express the analytical formulas for output ergotropy and output non-equilibrium free energy.
Here we consider three examples of phase-insensitive one-mode BGCs~\cite{caruso_onemode}:
the noisy lossy channel $\loss_{\mu,N}$, defined by the matrices
\begin{eqnarray}
X(\loss_{\eta,N}) = \sqrt{\eta} I_2  \;,  \qquad Y(\loss_{\eta,N}) = (1 - \eta)(2N+1) I_2 \; ;
\label{matrici_loss}
\end{eqnarray}
 the noisy amplification channel $\amp_{\mu,N}$, definded as
\begin{eqnarray}
X(\amp_{\mu,N}) = \sqrt{\mu} I_2   \;,  \qquad  Y(\amp_{\mu,N}) = (\mu - 1)(2N+1) I_2 \; ;
\label{matrici_amp}
\end{eqnarray}
and the additive noise channel $\noise$, defined as
\begin{eqnarray}
X(\noise) =  I_2   \;,  \qquad  Y(\noise) = 2N I_2 \; .
\label{matrici_noise}
\end{eqnarray}
Here $\eta \in [0,1]$, $\mu \in [1,\infty)$, and $N \in \mathbb{R}^+$ is 
the environmental noise term~\cite{caruso_onemode}.

As we have proven in the main section, the optimal state will always be a coherent state $\cohstate$.
For all of these three channel, the output covariance matrix $\sigma(\Phi(\cohstate))$ will be a multiple of the identity:
\begin{eqnarray} \nonumber
\sigma(\loss_{\eta,N}(\cohstate)) &=& (2N(1-\eta)+1)I_2\;, \\ \nonumber
\sigma(\amp_{\mu,N}(\cohstate)) &=& (2\mu(N+1)-2N-1)I_2 \;,\\ 
\sigma(\noise(\cohstate)) &=& (2N+1)I_2\;.
\end{eqnarray}
Now observe that when the covariance matrix is a multiple of the identity, i.e. $\sigma(\gstate) = \alpha I_2$, the ergotropy~(\ref{ergo_gaussiani_1modo}) can be simplified as
\begin{eqnarray}
\label{ergo_gaussiani_1modo_alphaId}
\ergo(\gstate) = \frac{m(\gstate)^2}{2} .
\end{eqnarray}
Therefore, the output ergotropy given an input coherent state with energy $E$ is:
\begin{eqnarray} \nonumber
\ergo(\loss_{\eta,N}(\cohstate)) = \eta E\;, \quad
\ergo(\amp_{\mu,N}(\cohstate)) = \mu E \;, \quad
\ergo(\noise(\cohstate)) = E\;.
\end{eqnarray}
By the property (12) of the main text these values also correspond to the maximum output values of the total ergotropy. 
The non-equilibrium free energy at fixed inverse temperature $\beta$ is instead given by
\begin{eqnarray} \nonumber
\freeen^{\beta}(\loss_{\eta,N}(\cohstate)) &=& \eta E + N(1-\eta)
- \tfrac{N(1-\eta)+1}{\beta}\log\left(N(1-\eta)+1\right) \nonumber \\
&&+ \tfrac{N(1-\eta)}{\beta}\log(N(1-\eta)) \;, \\ \nonumber 
\freeen^{\beta}(\amp_{\mu,N}(\cohstate)) &=& \mu E + (N+1)(\mu-1) 
- \tfrac{\mu(N+1)-N}{\beta} \log\left(\mu(N+1)-N\right) \\ \nonumber &&+ \tfrac{\mu(N+1)-N-1}{\beta}\log\left(\mu(N+1)-N-1\right)\;,\\ 
\freeen^{\beta}(\noise(\cohstate)) &=& E + N
- \tfrac{N+1}{\beta} \log\left(N+1\right) + \tfrac{N}{\beta} \log\left(N\right).
\end{eqnarray}

\subsection{General one-mode BGCs}
In the rest of this section we want to optimize on general BGC.  Since the problem becomes much more difficult we now restrict our analysis to one-mode BGCs and we only optimize on input Gaussian states. That is, we seek to find the Gaussian state $\gstate$, with $\en(\gstate)=E$, which maximises the output ergotropy $\ergo(\Phi(\gstate))$ for a general one-mode BGC $\Phi$.

Lemma~\ref{lemma:convexity_ergo} implies that, in order to optimize on the whole convex hull of Gaussian states, it is sufficient to consider pure Gaussian states, i.e. 
$\det(\sigma(\gstate)) = 1$.
This condition means that the covariance matrix of a pure one-mode gaussian state $\gstate$ must have eigenvalues $z$ and $z^{-1}$, where $z > 1$ is the degree of squeezing of the state; i.e., $\Tr[\sigma(\gstate)] = z + z^{-1}$. Using~(\ref{def:energy}), we can see that if $\gstate$, is a gaussian state of energy $E$, then
\begin{equation} 
\label{m_unmodo}
m(\gstate)^2 = m^2(z) := \frac{1}{2} \left( E + 1 - \frac{z + z^{-1}}{2} \right)\;.
\end{equation}
This equation fixes the module of the mean value vector $m(\gstate)$. Its optimal direction will be fixed by equation~(\ref{direzioneottimale}) below.
Condition~(\ref{m_unmodo}) limits the squeezing parameter $z$ to a maximum value of
\begin{equation}
z_{\max}(E) = 2E - 1 + \sqrt{(2E - 1)^2 - 1}\;.
\end{equation}
In order to optimize the ergotropy, we have to choose the direction vector $m(\gstate)$ (whose module is fixed by~(\ref{m_unmodo})) in order to satisfy 
\begin{eqnarray}
\label{direzioneottimale}
m^2(\Phi(\gstate)) = |Xm(\gstate)|^2 = \Lambda_1 m^2(\gstate) \; .
\end{eqnarray}
Using~(\ref{ergo_gaussiani_1modo}) and~(\ref{direzioneottimale}), we can express the output ergotropy as
\begin{eqnarray}
\label{output_ergotropy}
\ergo(\Phi(\gstate)) = \frac{\Lambda_1}{4} \left( E + 1 - \frac{z + z^{-1}}{2} \right) + \frac{\Tr[X \sigma(\gstate) X^T + Y]}{4}\;.
\end{eqnarray} 
We can parametrize the covariance matrix $\sigma(\gstate)$ of a pure state in terms of two parameters $z^\ast \in [1, \infty)$ and $\theta^\ast \in [0, 2\pi)$ -- see Eq.~(\ref{parametrizzazione_covarianza}) below.

The energy value $E$ appears in~(\ref{output_ergotropy}) only as a constant, and is therefore ininfluent in determing the optimal input. This means that, there will be an optimal covariance matrix $\sigma^\ast$, characterized by parameters $z^\ast \in [1, \infty]$ and $\theta^\ast \in [0, 2\pi)$, which does not depend on the energy $E$.
If $z^\ast < z_{max}(E)$, then the optimal state will be the one with covariance $\sigma(\gstate) = \sigma^\ast$, and whose mean value $m(\gstate)$ satisfies Eqs.~(\ref{m_unmodo}) and~(\ref{direzioneottimale}). If instead $z^\ast > z_{max}(E)$, the optimal input state will have $m(\gstate) = 0$ and squeezing $z_{max}$. In particular, if $z^\ast = \infty$, the optimal input will always be the maximally squeezed state available with energy $E$.
The plots in the main text are obtained by numerically finding the optimal $\sigma^\ast$, and then applying the rules described above.

It is worth to notice that, for energies $E \gg z^\ast$, the optimal input state $\gstate^\ast$ will have almost all of its energy stored in the mean value $m(\gstate^\ast)$. This implies that almost all the energy of the output state will be the (extractable) energy stored in its mean value $m(\Phi(\gstate^\ast)) = \sqrt{\Lambda_1} m(\gstate^\ast)$. Therefore, the asymptotic behaviour for large energies will be
\begin{eqnarray}
\max_{\en(\gstate) = E} \ergo(\Phi(\gstate)) \approx \Lambda_1 E \; ,
\end{eqnarray}
with 
\begin{eqnarray}
\max_{\en(\gstate) = E} \ergo(\Phi(\gstate)) / \en{\Phi(\gstate)} \approx 1 \; .
\end{eqnarray}

\subsection{Calculation of the optimal input covariance $\sigma^\ast$}

If we define
\begin{equation}
f_+ (z) := \frac{1}{2} \left( z + z^{-1} \right) \; ; \quad 
f_- (z) := \frac{1}{2} \left( z - z^{-1} \right)\;,
\end{equation}
then we can parametrize the covariance matrix of any gaussian pure state $\gstate$ as
\begin{equation}
\label{parametrizzazione_covarianza}
\sigma(\gstate) = f_+ (z) I_{2} +  f_- (z) \sin(\theta) \sigmax + f_- (z) \cos(\theta) \sigmaz\;, 
\end{equation}
where $z \in \left[ 1, z_{\max}(E) \right]$ is the degree of squeezing of $\gstate$,  $\theta \in \left[ 0 , 2\pi \right)$ is the squeezing direction, and 
$\sigmaz$ and $\sigmax$ are Pauli matrices.

\begin{figure} 
	\centering
	\includegraphics[width=\columnwidth]{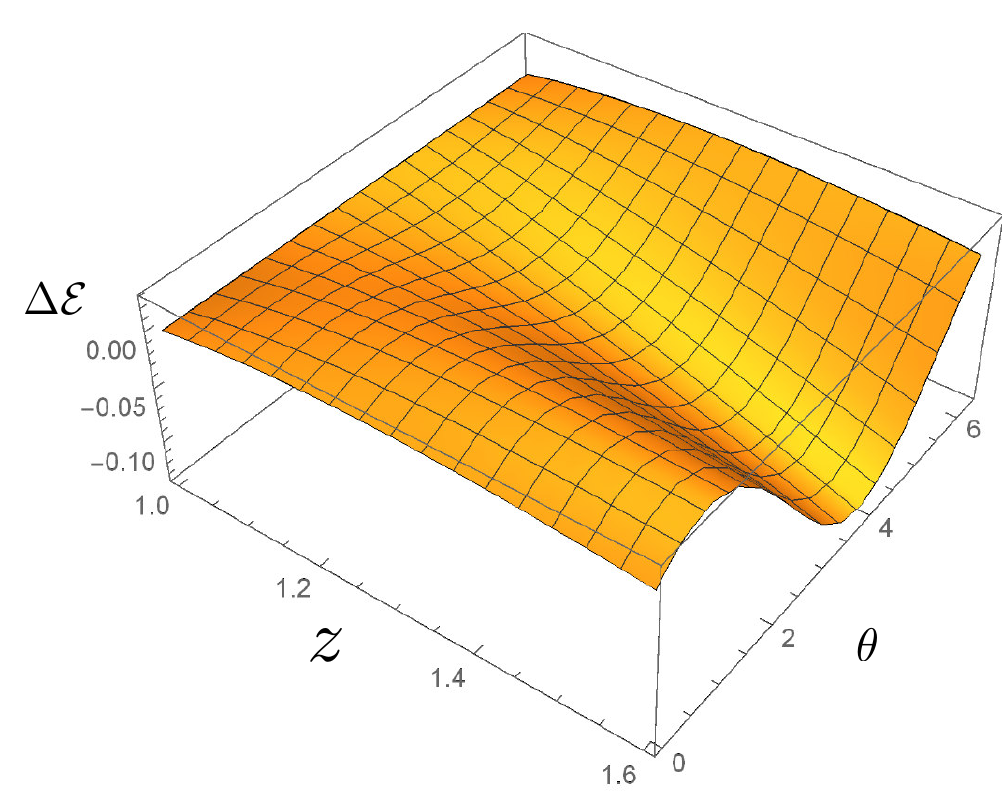}
	\caption{Difference $\Delta\ergo := \ergo(\Phi(\gstate)) -\ergo(\Phi(\cohstate))$ between the ergotropy $\ergo(\Phi(\gstate))$ of the output a gaussian state $\gstate$ squeezed of $z$ along the direction $\theta$, and the ergotropy $\ergo(\Phi(\cohstate))$ of a coherent state $\cohstate$ with the same energy as $\gstate$ and with optimal $m(\cohstate)$, for a one-mode BCG with $\Lambda_1 = 2$, $\Lambda_2 = 1.3$, $y_I = 1$, $y_X = 0.4$ and $y_Z = 0$.}
	\label{fig:superficie_ergotropia}
\end{figure}

In the one-mode setting we have a useful group theoretic property, namely the fact that $SO(2) \subset Sp(2,\mathbb{R})$; also, we are interested in functionals that are quadratic in $X$, so using the singular value decomposition without loss of generality we can assume
\begin{equation} 
X = 
\begin{bmatrix}
\sqrt{\Lambda_1} & 0 \\
0 & \sqrt{\Lambda_2}
\end{bmatrix}\;,
\end{equation}
with $\Lambda_1 \geq \Lambda_2$. Hence if we let
\begin{equation}
Y = y_I I_{2} + y_x \sigmax + y_z \sigmaz\;, 
\end{equation}
then, 
the covariance matrix $\sigma(\Phi(\gstate))$ of the output state $\Phi(\gstate)$ is:
\begin{eqnarray}
\sigma(\Phi(\gstate)) &=& X\sigma(\gstate) X^T + Y  \nonumber \\
&=& \left( \frac{\Lambda_1 + \Lambda_2}{2} f_+ (z) + \frac{\Lambda_1 - \Lambda_2}{2} f_- (z) \cos(\theta)     + y_I \right)  I_2   \nonumber \\
&&+\left( \sqrt{\Lambda_1\Lambda_2} f_-(z) \sin(\theta)    + y_x \right)  \sigmax  \nonumber \\
&&+ \left( \frac{\Lambda_1 - \Lambda_2}{2} f_+ (z) + \frac{\Lambda_1 + \Lambda_2}{2} f_- (z) \cos(\theta)     + y_z \right)  \sigmaz  \; .
\end{eqnarray}
So the eigenvalues of $\sigma(\Phi(\gstate))$ are:
\begin{eqnarray}
\label{autovalori_covoutput}
\lambda_{1,2}(z, \theta) &=& 
\left( \tfrac{\Lambda_1 + \Lambda_2}{2} f_+ (z) + \tfrac{\Lambda_1 - \Lambda_2}{2} f_- (z) \cos(\theta)     + y_I \right) \nonumber \\
&& \pm  \left[ \left( \sqrt{\Lambda_1\Lambda_2} f_-(z) \sin(\theta)    
+ y_x \right)^2 \right. \\
&& \; \left.  + \left( \tfrac{\Lambda_1 - \Lambda_2}{2} f_+ (z) + \tfrac{\Lambda_1 + \Lambda_2}{2} f_- (z) \cos(\theta)     + y_z \right)^2 \right]^{1/2} \nonumber\;,
\end{eqnarray}
and we can write the ergotropy~(\ref{ergo_gaussiani_1modo}) as
\begin{eqnarray} \label{ergo_autovalori}
\ergoparam(z, \theta) &=& 
\frac{\lambda_1(z, \theta) + \lambda_2(z, \theta)}{4} - \frac{\sqrt{\lambda_1(z, \theta) \lambda_2(z, \theta)}}{2} + \frac{\widetilde{m}^2(z)}{2}  \nonumber \\
&=& \frac{\left( \sqrt{\lambda_1(z, \theta)} - \sqrt{\lambda_2(z, \theta)} \right)^2 }{4} + \frac{\widetilde{m}^2(z)}{2}\;, 
\label{ergo_alphatheta}
\end{eqnarray}
here $\ergoparam(z,\theta) := \ergo(\Phi(\gstate))$ and $\widetilde{m}(z) := m(\Phi(\gstate))$, where $z$ and $\theta$ are the parameters of the gaussian state $\gstate$. 
Figure \ref{fig:superficie_ergotropia} shows a plot of $\ergoparam(z, \theta) - \ergoparam(1, 0) = \ergoparam(z, \theta) -\frac{\widetilde{m}^2(1)}{2} $, as a function of $z$ and $\theta$, for an example Gaussian channel.

\begin{figure}
	\centering
	\includegraphics[width=\linewidth]{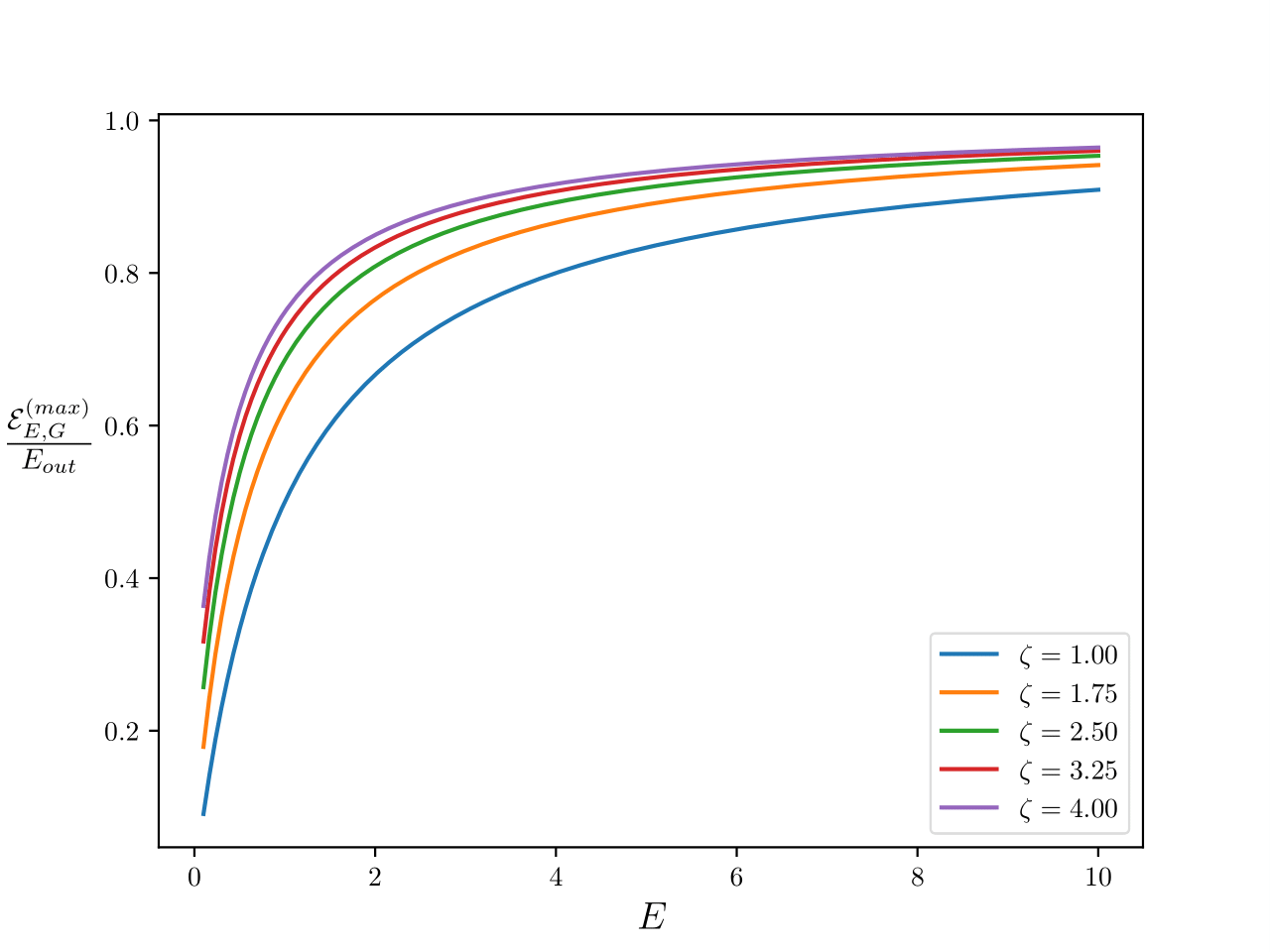}
	\caption{Ratio $\mathcal{E}^{(max)}_{E, G} / E_{out}$ between the maximum output ergotropy $\mathcal{E}^{(max)}_{E, G}$ and the output energy $E_{out}$, for the channel $\Gamma_{\eta, \zeta}$ composed by a squeezing transformation $\Sigma_\zeta$ followed by a quantum-limited attenuator $\loss_{\eta,0}$, with $\eta=0.5$, for various values of the squeezing parameter $\zeta$. }
	\label{fig:att_sq}
\end{figure}

\begin{figure}
	\centering
	\includegraphics[width=\columnwidth]{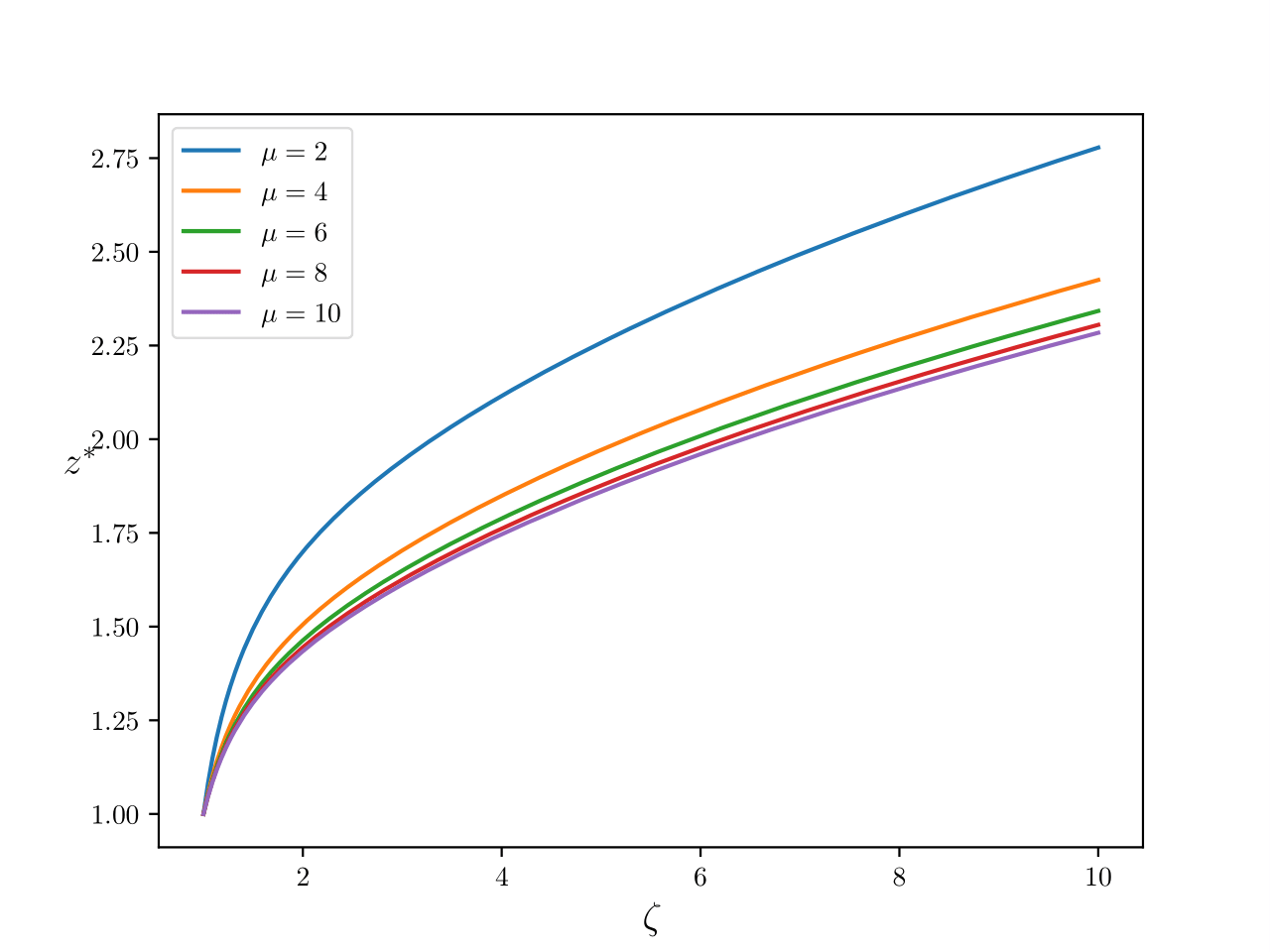}
	\caption{Optimal squeezing $z^*$ for some channels $\Theta_{\mu, \zeta}$ composed by a unitary squeezing of parameter $\zeta$ followed by a quantum-limited amplifier of gain $\mu$ (as defined by the matrices in Eq.~(\ref{matrici_theta2})). }
	\label{fig:opt_z}
\end{figure}

For any given channel, characterized by the matrices $X$ and $Y$, and a given input energy $E$, the function~(\ref{ergo_autovalori}) can be maximised numerically to find the optimal Gaussian state.
In figure \ref{fig:att_sq} we show the behavior of optimal ergotropy of different phase-sensitive attenuators $\Gamma_{\eta,\zeta} =  \loss_{\eta,0} \circ \Sigma_\zeta$ by varying the input energy $E$; here $\loss_{\eta,0}$ is a quantum-limited attenuator charachterized by the matrices~(\ref{matrici_loss}),
while $\Sigma_\zeta$ is a diagonal unitary squeezing with parameter $\zeta$, defined by
\begin{eqnarray}
X(\Sigma_\zeta) = 
\begin{pmatrix}
\sqrt{\zeta} & 0 \\
0 & \sqrt{\zeta^{-1}}
\end{pmatrix}
 \;, \qquad Y(\Sigma_\zeta) = 0_2 \; .
\label{matrici_squeezing}
\end{eqnarray}
Therefore, the composite channel $\Gamma_{\eta,\zeta}$ is characterized by the matrices
\begin{eqnarray}
X(\Gamma_{\eta,\zeta}) &=&  X(\loss_{\eta,0}) X(\Sigma_\zeta) = 
\sqrt{\eta}\begin{pmatrix}
\sqrt{\zeta} & 0 \\
0 & \sqrt{\zeta^{-1}}
\end{pmatrix}\;,
\nonumber \\ 
Y(\Gamma_{\eta,\zeta}) &=& X(\loss_{\eta,0}) Y(\Sigma_\zeta) X^\dagger(\loss_{\eta,0}) + Y(\loss_{\eta,0}) =
(1-\eta)\begin{pmatrix}
1 & 0 \\
0 & 1
\end{pmatrix}  \; .
\label{matrici_theta}
\end{eqnarray}
In Figure \ref{fig:opt_z} we plot the optimal squeezing parameter $z^*$ for some channels in the form $\Theta_{\mu,\zeta} = \amp_{\mu,0} \circ \Sigma_\zeta $, where $\Sigma_\zeta$ is a squeezing channel of squeezing parameter $\zeta$ (see~(\ref{matrici_squeezing})), and $\amp_{\mu,0}$ is a quantum-limited amplifier of gain $\mu$, defined by setting $N=0$ in~(\ref{matrici_amp}),
so that
\begin{eqnarray}
X(\Theta_{\mu,\zeta}) &=&  X(\amp_{\mu,0}) X(\Sigma_\zeta) = 
\sqrt{\mu}\begin{pmatrix}
\sqrt{\zeta} & 0 \\
0 & \sqrt{\zeta^{-1}}
\end{pmatrix} \;, 
\nonumber \\
Y(\Theta_{\mu,\zeta}) &=& X(\amp_{\mu,0}) Y(\Sigma_\zeta) X^\dagger(\amp_{\mu,0}) + Y(\amp_{\mu,0}) =
(\mu-1)\begin{pmatrix}
1 & 0 \\
0 & 1
\end{pmatrix}  \; 
\label{matrici_theta2}.
\end{eqnarray}

\begin{figure}
	\centering
	\includegraphics[width=\linewidth]{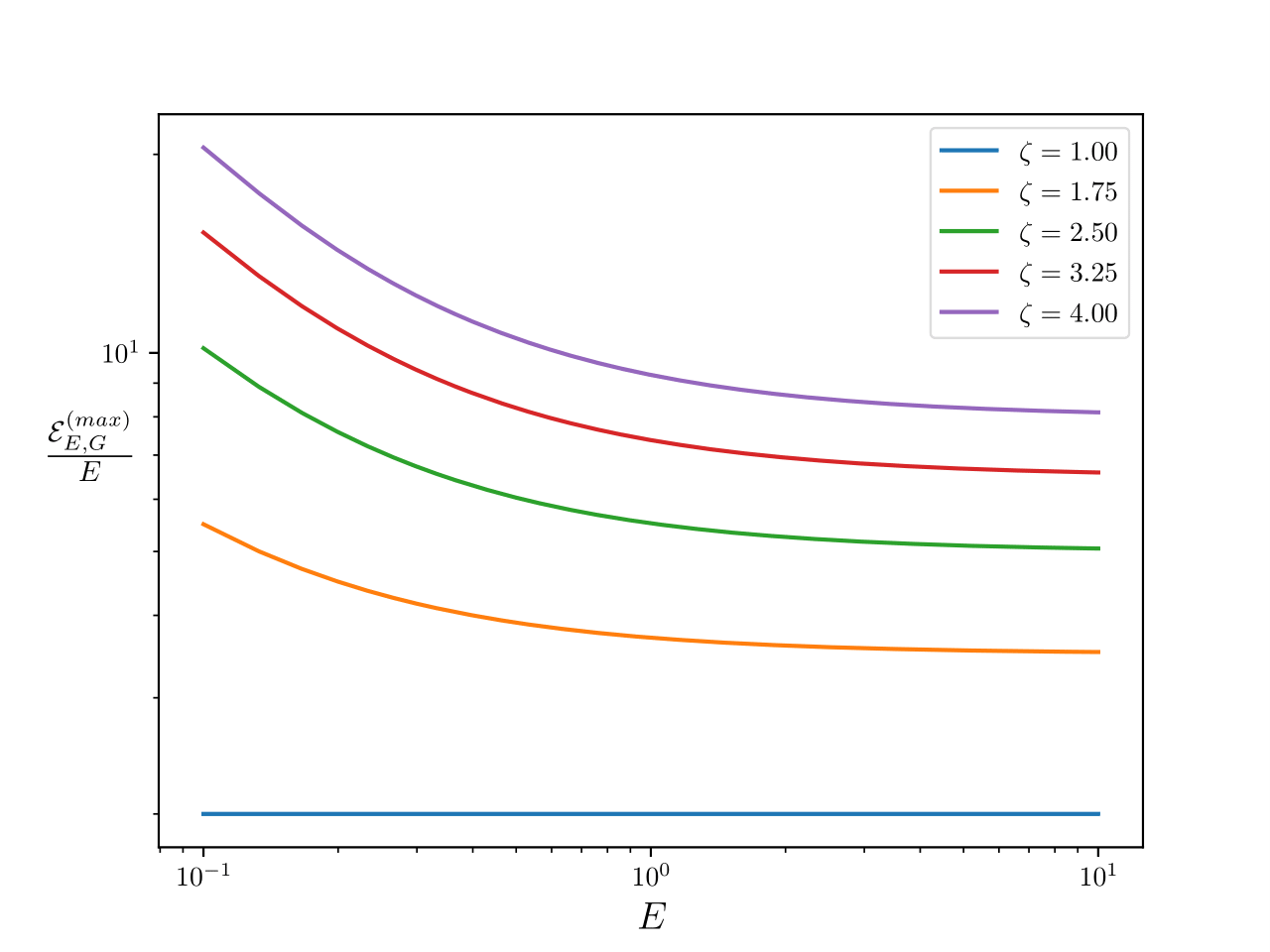}
	\caption{Ratio $\mathcal{E}^{(max)}_{E, G} / E$ between the maximum output ergotropy $\mathcal{E}^{(max)}_{E, G}$ and the input energy $E$, for the channel $\Theta_{\mu,\zeta}$ composed by a squeezing transformation $\Sigma_z$, followed by a quantum-limited amplifier $\amp_{\mu,0}$, with $\mu=2$  (see~(\ref{matrici_theta})), for various values of the squeezing parameter $\zeta$. 
	}
	\label{fig:att_sq2}
\end{figure}

\subsection{Special cases}
\label{sec:special_cases}
We now focus on the special case in which the matrix $Y$ is diagonal in the base of the squeezing, i.e. $y_x=0$.
Replacing~(\ref{autovalori_covoutput}) into~(\ref{ergo_alphatheta}) and deriving we can see that, when $y_x = 0$, $\frac{\partial \ergoparam(z, \theta)}{\partial \theta} = 0$, for any fixed value of the squeezing parameter $z$, the maximum and the minumum value of $\ergoparam(z, \theta)$ are reached when $\theta = 0$ and $\theta = \pi$, so the covariance matrix must be diagonal in the squeezing basis and its eigenvector with the largest eigenavalue must be parallel to the one of the squeezing matrix. 
The optimal squeezing $z^\ast$ can then be found by setting $\frac{\partial \ergoparam(z, \theta)}{\partial z} = 0$, i.e. by solving the eight-grade equation
\begin{eqnarray} 
\label{eq: derivatona}
\frac{\partial \ergoparam(z, 0)}{\partial z} 
= \tfrac{1}{4z}  \left[ 
\tfrac{ -2z^2 \Lambda_1 y_I + 2\Lambda_2 y_I + 2z^2\Lambda_1 y_z + 2\Lambda_2 y_z }{\sqrt{\left( z^2\Lambda_1 + \Lambda_2 + 2z y_I \right)^2 - \left( z^2\Lambda_1 - \Lambda_2 + 2z y_z \right)}} +  \tfrac{\Lambda_1 - \Lambda_2}{z}  \right] = 0 \;. 
\end{eqnarray}
This optimal squeezing can only be reached if $z^\ast \leq z_{max}(E)$; otherwise the optimal allowed gaussian state will be the one with maximal squeezing $z_{max}(E)$. The maximum ergotropy will then be given by
\begin{eqnarray}
\max \ergo(\Phi(\gstate)) 
= \begin{dcases}
\ergoparam(\min(z^\ast, z_{max}(E)) , 0) & \text { if } z^\ast \geq 1 \; ; \\
\ergoparam(\min(\left( z^\ast \right)^{-1}, z_{max}(E)) , \pi) & \text { if } 0 < z^\ast \leq 1 \; .
\end{dcases} 
\end{eqnarray} 
In the above scenario it can be interesting to consider the stronger condition $y_Z = y_X = 0$; in this case $Y$ is proportional to the identity matrix and Eq. (\ref{eq: derivatona}) can be greatly simplified to a fourth-grade equation: 
\begin{eqnarray}
\frac{\partial \ergoparam(z, \theta)}{\partial z} =
- \frac{1}{4 z} \left[ y_I \tfrac{\Lambda_2 / z^2 - \Lambda_1}{\sqrt{\left( y_I + \Lambda_1z \right) \left( y_I + \Lambda_2 / z \right)}} + \frac{\Lambda_1 - \Lambda_2}{z}  \right] = 0 \; . \nonumber \\
\end{eqnarray}
